\documentclass[twocolumn,floatfix,pra,aps,showpacs]{revtex4}
\usepackage{epsfig,graphicx}
\usepackage{flafter}
\usepackage{amsmath}
\usepackage{color}
\usepackage{braket}

\begin{document}

\title{Scaling solutions of the two fluid hydrodynamic equations in a  harmonically trapped  gas at unitarity }
\author{ Yan-Hua Hou$^{1}$, Lev P. Pitaevskii$^{1,2}$, Sandro Stringari$^{1}$}
\affiliation{1 Dipartimento di Fisica, Universit\`{a} di Trento
and INO-CNR BEC Center, I-38123 Povo, Italy\\
2 Kapitza Institute for Physical Problems, Russian Academy of Science, Kosygina
2, 119334 Moscow, Russia}

\date{\today}
\begin{abstract}
We prove that the two fluid Landau hydrodynamic equations, when applied to a  gas interacting with infinite scattering length (unitary gas) in the presence  of harmonic trapping, admit exact scaling solutions of mixed compressional and surface nature. These solutions are characterized by a linear dependence of the velocity field on the spatial coordinates and a temperature independent  frequency which is  calculated in terms of the parameters of the trap.   Our results are  derived  in the regime of  small amplitude oscillations and  hold  both below and above the superfluid phase transition. They apply to  isotropic as well as to deformed configurations, thereby   providing a generalization of  Castin's theorem (Y. Castin, C. R. Phys. \textbf{5}, 407 (2004)) holding for isotropic trapping. Our predictions agree with the experimental findings in resonantly interacting atomic Fermi gases. The breathing scaling solution, in the presence of isotropic trapping, is  also used to prove the vanishing of two bulk viscosity coefficients in the superfluid phase.

\end{abstract}

\maketitle


The thermodynamic behavior of Fermi gases interacting with infinite
scattering length (hereafter called unitary Fermi gases) are known to
exhibit peculiar universal properties. This follows from the fact that in
the unitary limit of uniform configurations the only available lengths are
the interparticle distance and the thermal wavelength. Analogously the
remaining energies are the thermal energy $k_BT$ and the Fermi energy $E_F=%
\frac{\hbar^{2}}{2m} (3\pi^{2}n)^{2/3}$, where $k_B$ is the Boltzmann constant and $n$ is the particle density. This peculiarity makes these
systems extremely interesting because their thermodynamic behavior takes a
universal character, suitable to explain several physical features of
quite different systems, like resonantly interacting atomic gases
and neutron matter. Systematic experimental \cite{Luo,Nascimbene,Horikoshi,MarkMartain} and theoretical (see for example \cite{Ho,GLS, BlochRMP} and refs therein)
efforts have been made to explore this universal thermodynamic behavior in
the superfluid as well as in the normal phase. The concept of unitarity is not restricted to the Fermi gas and applies to Bose gases as well \cite{LHO,Salomon,Stoof}, although three-body recombinations cause a fast instability of these systems, making their experimental investigation  problematic especially at low temperature.

The unitary Fermi gas is known
to exhibit unique features also from the dynamic point of view. In
particular Castin  \cite{castin} has shown that exact scaling solutions are
available if the system is trapped by an isotropic three-dimensional
harmonic potential. A remarkable example is the occurrence of an undamped
radial breathing mode oscillating at the frequency $2\omega_{ho}$, where $%
\omega_{ho}$ is the oscillator frequency of the harmonic potential. This
result is remarkable because it concerns a strongly interacting system and
its validity is not restricted to small amplitude oscillations. Furthermore
it holds exactly at all temperatures irrespective of the collisional regime
and the value of the mean free path. It can be regarded as the strongly
interacting and quantum analog of the most famous classical result derived
by Boltzmann for an ideal gas trapped by an isotropic harmonic potential~\cite{Cerc88}.

Similarly to the case of the Boltzmann gas, the universality of the scaling
oscillation breaks down in the presence of a deformed harmonic trap. In this case no general exact result is available, unless one considers the collisional hydrodynamic regime characterized by
the occurrence of fast collisions. This is the case considered in the
present paper where we prove the existence of a class of scaling solutions
describing small amplitude oscillations of the gas around  equilibrium,
characterized by a temperature independent value of the
collective frequency. These solutions have been already identified in the
literature at zero temperature (see \cite{GLS} and references therein). Their existence is proven here at all temperatures, both below and above the critical temperature for
superfluidity. For simplicity we consider an axially symmetric trapping
potential

\begin{equation}
V_{ext}= \frac{1}{2}m\omega^{2}_{\perp}\mathbf{r}^{2}_{\perp}+\frac{1}{2}%
m\omega^{2}_{z}\mathbf{z}^{2}  \label{VHO}
\end{equation}
where  $\omega_{\perp}$ and $\omega_z$ are, respectively, the radial frequencies while  $\mathbf{r}_{\perp}=x \hat{i}+y \hat{j}$,  but our proof can be easily generalized to the more general case of tri-axial trapping.

We start from the two fluid hydrodynamic equations
\begin{eqnarray}  \label{A1}
m\partial_{t}n+\nabla \cdot\mathbf{j} =0
\end{eqnarray}
\begin{eqnarray}  \label{A2}
\partial_{t}s+\nabla\cdot(s\mathbf{v}_n)=0
\end{eqnarray}
\begin{eqnarray}  \label{A3}
m\partial_{t}\mathbf{v}_{s}=-\nabla (\mu+V_{ext})
\end{eqnarray}
\begin{eqnarray}  \label{A4}
\partial_{t}\mathbf{j}=-\nabla P-n\nabla V_{ext} \; ,
\end{eqnarray}
first derived by Landau \cite{Landau41}, which describe the dynamic behavior
of a gas in its superfluid phase. Equation (\ref{A1}) is the equation of
continuity where $n$ is the density of the fluid and $\mathbf{j} =m(n_{n}\mathbf{v}%
_{n}+n_{s}\mathbf{v}_{s})$ is the current density where we have introduced the
superfluid ($n_{s}$) and normal densities ($n_{n}$) and the corresponding
velocity fields $\mathbf{v}_{n}$ and $\mathbf{v}_{s}$. Equation (\ref{A2}) shows
that the transport of entropy ($s$ is the entropy density) is carried by
the normal component of the fluid. Equation (\ref{A3}) instead fixes the equation for the
superfluid velocity,  governed by the chemical potential ($\mu=\mu(n,T)$ is
the local value of the chemical potential, determined by the equation of
state of uniform matter). Finally Eq. (\ref{A4}) is the Euler equation for
the current fixed by the gradient of the pressure $P$ and of the external
potential $V_{ext}$. Above the superfluid critical temperature the
superfluid density vanishes and Eqs. (\ref{A1}-\ref{A2}) and (\ref{A4})
reduce to the usual hydrodynamic equations of a normal fluid.

In the following we prove the existence of exact solutions of the
hydrodynamic equations, corresponding to the first sound like ansatz $\mathbf{v}%
_{n}=\mathbf{v}_s\equiv \mathbf{v}$ for the velocity field of the normal and
superfluid components with:
\begin{eqnarray}  \label{A5}
 \mathbf{v}= \beta(t)\mathbf{r}_{\perp}+\delta(t)\mathbf{z}\; .
\end{eqnarray}
Above $T_c$, where the superfluid density is absent, the same anstaz applies
to the velocity field  of the fluid. In Eq. (\ref{A5}) we have
considered a velocity flow with symmetric behavior in the $x$ and $y$
directions, corresponding to excitations carrying angular momentum along the z-th direction $\ell_z=0$%
. The generalization of the formalism to scaling excitations carrying
angular momentum $\ell_z=\pm 1$ and $\ell_z = \pm 2$ is straightforward.

The choice (\ref{A5}) is accompanied by the following scaling
transformations for the density
\begin{eqnarray}  \label{A6}
n(\mathbf{r},t)=e^{2\alpha(t)+\gamma(t)}n_{0}(\mathbf{r}^\prime)
\end{eqnarray}
and for the entropy density
\begin{eqnarray}  \label{A7}
s(\mathbf{r},t)=e^{2\alpha(t)+\gamma(t)}s_{0}(\mathbf{r}^\prime)
\end{eqnarray}
where $\mathbf{r}^\prime\equiv
(e^{\alpha(t)}x,e^{\alpha(t)}y,e^{\gamma(t)}z) $ is the scaled spatial
variable and $n_0$ and $s_0$ are the particle density and entropy density
calculated at  equilibrium. The prefactor $e^{2\alpha(t)+%
\gamma(t)}$ in the above equations ensures the normalization condition for
the density and the conservation of total entropy. The local entropy per particle $s(\mathbf{r})/n(\mathbf{r})$ is also conserved. At unitary the entropy
density can be written in the form $s(n, T)= n f_e(T/T_F(n))$ where
$T_F=E_F/k_{B}$ is the Fermi
temperature and $f_e$ is a universal function that can be derived by the
knowledge of the equation of state (see for example \cite{Hou}), but whose
explicit form is irrelevant for the proof of our theorem. The ansatz (\ref{A7}) then requires that  the ratio $T/T_F(n)$ should be conserved by the scaling transformation. This implies that  the temperature should exhibit
the position independent scaling law
\begin{equation}  \label{Tt}
T(t)= e^{(2\alpha(t)+\gamma(t))2/3}T_0
\end{equation}
where $T_0$ is the temperature of the gas at equilibrium. Finally the
chemical potential, which due to dimensionality arguments can be written in
the form $\mu= T_F(n)f_\mu(T/T_F(n))$ where $f_\mu$ is a dimensionless
function, exhibits the following scaling
behavior
\begin{eqnarray}  \label{A8}
\mu(\mathbf{r},t) =e^{(2\alpha(t)+\gamma(t))2/3}\mu_{0}(n_{0}(\mathbf{r}%
^\prime))
\end{eqnarray}
with $\mu_0$ calculated at the equilibrium temperature $T_0$. The fact that
temperature fluctuations associated with the scaling solutions are uniform
in space represents a peculiar feature of these collective oscillations. It
implies, in particular, that as a consequence of the thermodynamic
relationship $\nabla P=s\nabla T+n\nabla\mu$ and of the ansatz (\ref{A5}),
the equations (\ref{A3}) and (\ref{A4}) for the superfluid velocity and for
the current are exactly equivalent.

We now prove that the above scaling ansatz actually  corresponds to an
exact solution of the hydrodynamic equations. From the equation of
continuity one finds the following relationship
\begin{eqnarray}  \label{A9}
[(2\dot{\alpha}+\dot{\gamma})+(2\beta+\delta)]n_0 & &\nonumber\\
+(\beta+\dot{\alpha}%
)\mathbf{r}_{\perp}\cdot\nabla_{\perp}n_0
& &\nonumber\\+(\delta+\dot{\gamma})z\nabla_{z}n_0=0
\end{eqnarray}
which implies the identities
\begin{eqnarray}  \label{A10}
\dot{\alpha}=-{\beta}, & &\dot{\gamma}=-{\delta}
\end{eqnarray}
The same conditions permit to satisfy the equation for the entropy density.

Since at equilibrium the chemical potential satisfies the condition $\nabla
\mu_0(\mathbf{r})=-m\omega_{\perp}^{2}\mathbf{r}_{\perp} -m\omega_{z}^{2}\mathbf{z}$
(see Eq.(\ref{A3})), the equation for the superfluid velocity (or, equivalently, the equation
for the total current) takes the simplified form
\begin{eqnarray}  \label{A11}
\dot{\beta}\mathbf{r}_{\perp}+\dot{\delta}\mathbf{z}&=&(e^{2\alpha+(2\alpha+%
\gamma)2/3}-1) \omega_{\perp}^{2}\mathbf{r}_{\perp}  \notag \\
& &+(e^{2\gamma+(2\alpha+\gamma)2/3}-1)\omega_{z}^{2}\mathbf{z}
\end{eqnarray}

By looking for time dependent solutions proportional to $e^{-i\omega t}$ and
expanding the exponentials of Eq.(\ref{A11}) up to terms linear in $\alpha$
and $\gamma$,  one finally derives the  coupled equations
\begin{eqnarray}  \label{Eq.12}
\omega^2 \alpha=\left(\frac{10}{3}\alpha+\frac{2}{3}\gamma\right)
\omega_\perp^2
\end{eqnarray}
and
\begin{eqnarray}  \label{Eq.13}
\omega^2 \gamma=\left(\frac{4}{3}\alpha+\frac{8}{3}\gamma\right) \omega^2_z
\end{eqnarray}
yielding the temperature independent result
\begin{eqnarray}  \label{Eq.14}
\omega^2= \left({\frac{5}{3}} +{\frac{4}{3}\lambda^2} \pm\frac{1}{3}\sqrt{16\lambda^4-32%
\lambda^2+25} \right) \omega^2_\perp
\end{eqnarray}
for the collective frequencies where $\lambda=\omega_{z}/\omega_{\perp}$ is  the aspect ratio of the trap.

For isotropic trapping ($\lambda=1$, i.e. $\omega_\perp =\omega_z \equiv
\omega_{ho}$) the corresponding modes are the uncoupled monopole (breathing)
mode with $\omega=2\omega_{ho}$ and the surface quadrupole mode with $\omega=%
\sqrt{2}\omega_{ho}$. For highly elongated  traps ($\lambda\ll 1$, i.e. $%
\omega_z \ll \omega_\perp$) the two solutions are instead $\omega=\sqrt{10/3}%
\omega_{\perp}$ and $\omega=\sqrt{12/5}\omega_{z}$, in agreement with the
results already derived in the literature at zero temperature (see for
example \cite{minguzzi, stringari2004}).

The above results provide a generalization of the theorem of \cite{castin},
where it was shown that for isotropic trapping the monopole breathing mode
of the unitary Fermi gas oscillates with frequency $\omega =2\omega _{ho}$,
independent of temperature. Differently from the case of \cite{castin} our
results  hold, however, only in the dissipationless
hydrodynamic regime and in the limit of small amplitude oscillations. The temperature independence of the frequency of the scaling solutions of the unitary Fermi gas has been already confirmed in experiments \cite{Altmeyer, joint}.

The scaling solutions discussed above  are characterized by temperature variations  and by an axial velocity field $v_z$  independent of the radial coordinates $x$ and $y$. These  are  the conditions required, in general, to apply the 1D hydrodynamic equations in highly elongated configurations (see \cite{Gianluca,Hou}) which have been recently successfully applied to describe the experimental results of \cite{joint}.  In general these 1D like conditions are ensured by the effective role of the thermal conductivity and of the viscosity  which cause  the absence of gradients in the radial direction and are favored by the presence of a tight radial confinement. In the case of the low frequency oscillations considered in the present work the absence of radial gradients is  automatically ensured by the form of the scaling transformation. This explains, in particular,  why the frequency   $\omega=\sqrt{12/5}\omega_{z}$  of the axial breathing mode, here derived  in a 3D framework in the highly elongated limit $\omega_z\ll \omega_\perp$, coincides with the predictions of the 1D hydrodynamic equations \cite{Hou}.

Let us also mention that the proof of the temperature independence of the frequency of the scaling solutions can be derived also for the $\ell_z=\pm 1$ and $\ell_z = \pm 2$ excitations  characterized by  velocity fields proportional to $\nabla[ z (x\pm iy)]$ and  $\nabla (x\pm iy)^2$,  respectively and yielding the values $\sqrt{\omega^2_\perp + \omega^2_z}$ and $\sqrt{2}\omega_\perp$ for the corresponding frequencies.  Differently from the $\ell_z=0$ solutions discussed above the results for the $\ell_z=\pm 1$ and $\ell_z=\pm 2$ are not restricted to the unitary Fermi gas, but simply  require the applicability of the hydrodynamic equations, being solutions characterized by divergency free velocity fields (surface excitations). They hold in particular for both Fermi and Bose gases in the presence of harmonic trapping.

Let us finally discuss a non trivial implication of the scaling solutions concerning the behavior of the bulk viscosity coefficients. According to Khalatnikov \cite{IMK} the entropy production
per unit volume associated with a hydrodynamic flow is $R/T$, where $R$ is the so called  dissipative function in the superfluid phase, defined by
\begin{eqnarray}
R &=&\frac{1}{2}\eta \left( \partial_k v_{ni}+
\partial_i v_{nk}-\frac{2}{3}\delta _{ik}\nabla\cdot%
\mathbf{v}_{n}\right) ^{2}  \notag \\
&&+2\zeta _{1}\nabla\cdot\mathbf{v}_{n}\nabla\cdot mn_{s}\left( \mathbf{v}_{s}-%
\mathbf{v}_{n}\right) +  \notag \\
&&+\zeta _{2}\left( \nabla\cdot\mathbf{v}_{n}\right) ^{2}+\zeta _{3}\left[
\nabla\cdot m n_{s}\left(\mathbf{v}_{s}-\mathbf{v}_{n}\right) \right] ^{2}
 \label{R}
\\
&&+\left( \kappa /T\right) \left( \nabla T\right) ^{2}\; .  \notag
\end{eqnarray}%

In the above equation $\eta $ is the shear viscosity, $\zeta _{1}$, $\zeta _{2}$ and $\zeta _{3}$ are the three bulk viscosity
coefficients  appearing in the superfluid phase, while $\kappa $ is the thermal conductivity. For the scaling modes discussed in the present paper $\nabla T=0$ and the velocity fields for the normal and superfluid components coincide.  Furthermore for the monopole breathing mode  in an isotropic trap one has
$\mathbf{v}_{n}=\mathbf{v}_{s}=\beta \mathbf{r}$ and the first term, proportional to the shear viscosity, identically vanishes.  It follows that in this
case only the term with $\zeta _{2}$ survives. However,  according to Castin theorem  \cite{castin} the dissipation
associated with the breathing oscillation must be  zero and we then
conclude that $\zeta _{2}=0$. On the other hand the positiveness of $R$ implies  that $\zeta_{1}^{2}\leq\zeta_{2}\zeta_{3}$ . Thus also $\zeta _{1}$ must vanish, the only surviving bulk viscosity coefficient being $\zeta _{3}$.  Above $T_{c}$, where one can introduce only one bulk viscosity term, the viscosity coefficient $\zeta$ should be also zero in order to ensure the absence of dissipation. The same results, concerning the value of the bulk viscosity coefficients, were previously derived by  Son \cite{Son2007} using different considerations. Actually, our derivation provides a simple foundation to Son's heuristic argumentation.

\acknowledgments
This
work has been supported by ERC through the QGBE grant.

\end{document}